\documentclass[usenatbib,letterpaper]{mn2e}

\usepackage{amsmath, amssymb}
\usepackage{mathrsfs}
\usepackage{array}
\usepackage{graphicx}
\usepackage{color}

\begin{document}

\newcommand{\be}{\begin{equation}}
\newcommand{\ee}{\end{equation}}
\newcommand{\pdf}{p}
\newcommand{\data}{{d}}
\newcommand{\Df}{{D}_f}
\newcommand{\Bf}{\mathcal{B}_{01}}
\newcommand{\lnBf}{\ln\Bf}
\newcommand{\mdl}{\mathcal{M}}
\newcommand{\lsim}{\,\raise 0.4ex\hbox{$<$}\kern -0.8em\lower
  0.62ex\hbox{$\sim$}\,} 
\newcommand{\gsim}{\,\raise 0.4ex\hbox{$>$}\kern -0.7em\lower
  0.62ex\hbox{$\sim$}\,} 

\newcommand{\params}{{\theta}}
\newcommand{\mean}{{\mu}}
\newcommand{\like}{L}
\newcommand{\lnlike}{\mathcal{L}}
\newcommand{\ML}{^*}
\newcommand{\dr}{\textrm{d}}
\newcommand{\ie}{i.e.}
\newcommand{\reion}{\text{re}}

\newcommand{\cd}{\cdot}
\newcommand{\cds}{\cdots}
\newcommand{\ip}{\int_0^{2\pi}}
\newcommand{\al}{\alpha}
\newcommand{\ba}{\beta}
\newcommand{\de}{\delta}
\newcommand{\De}{\Delta}
\newcommand{\ep}{\epsilon}
\newcommand{\Ga}{\Gamma}
\newcommand{\ka}{\tau}
\newcommand{\io}{\iota}
\newcommand{\La}{\Lambda}
\newcommand{\Om}{\Omega}
\newcommand{\om}{\omega}
\newcommand{\si}{\sigma}
\newcommand{\Si}{\Sigma}
\newcommand{\te}{\theta}
\newcommand{\ze}{\zeta}
\newcommand{\vth}{\ensuremath{\vartheta}}
\newcommand{\vph}{\ensuremath{\varphi}}
\newcommand{\MM}{\mbox{$\cal M$}}
\newcommand{\tr}{\mbox{tr}}
\newcommand{\hor}{\mbox{hor}}
\newcommand{\grad}{\mbox{grad}}
\newcommand{\cx}{\ensuremath{\mathbf{\nabla}}}
\newcommand{\lap}{\triangle}
\newcommand{\arctg}{\mbox{arctg}}
\newcommand{\bm}[1]{\mbox{\boldmath $#1$}}
\newcommand{\eff}{{\rm eff}}
\newcommand{\tto}{\Rightarrow}
\newcommand{\lag}{\langle}
\newcommand{\rag}{\rangle}
\newcommand{\fiso}{f_{\text{iso}}}
\newcommand{\Afiso}{\vert f_{\text{iso}}\vert}
\newcommand{\norm}[3]{{N_{#1,#2}(#3)}}
\newcommand{\fexp}{\boldsymbol{e}}
\newcommand{\KL}{D_{KL}}
\newcommand{\ns}{n_S}
\newcommand{\omk}{\Om_\kappa}
\newcommand{\LL}{{\cal L}}
\newcommand{\bth}{\bar{\theta}}
\newcommand{\thb}{\bth}
\newcommand{\thnot}{\theta_0}
\newcommand{\id}{{\mathbb I}}
\newcommand{\ev}{{\cal E}}
\newcommand{\Lnotf}{{\LL_0^f}}
\newcommand{\xf}{x_f}
\newcommand{\Lf}{{\mathbb L}}
\newcommand{\Af}{{\mathbb A}}
\newcommand{\Ff}{{\mathbb F}}
\newcommand{\thnotf}{\theta_0^f}
\newcommand{\Psinotf}{\Psi_0^f}
\newcommand{\fid}{{\params}_\star}
\newcommand{\thbf}{\bth^f}
\newcommand{\Yf}{Y_f}
\newcommand{\Xf}{X_f}
\newcommand{\Mf}{{\mathbb M}}
\newcommand{\fom}{{\rm FoM}}
\newcommand{\dB}{\Delta \ln B}
\newcommand{\Uf}{{\mathcal U}}
\newcommand{\EU}{{\mathcal{EU}}}
\newcommand{\Dec}{{\mathscr{D}}}
\newcommand{\Expu}{{\mathscr{E}}}
\newcommand{\tp}{\theta_\pi}
\newcommand{\Fff}{\mathcal{F}}
\newcommand{\rt}[1]{{\bf RT: #1}}
\newcommand{\mk}[1]{{\tt MK: #1}}

\title[Designing Decisive Detections]{Designing Decisive Detections}

\author[Trotta, Kunz \& Liddle]{Roberto Trotta,$^{1,2}$ Martin Kunz$^{3,4}$ and Andrew R.~Liddle$^4$\\
$^1$Astrophysics Group, Imperial College London, Blackett Laboratory,
  Prince Consort Road, London SW7 2AZ, UK \\ 
$^2$Oxford University, Astrophysics Department, Denys Wilkinson
Building, Keble Road, Oxford, OX1 3RH, UK \\  
$^3$University of Geneva, D\'epartment de Physique Th\'eorique, Quai E. Ansermet 24, 1205 Geneva, Switzerland \\ 
$^4$Astronomy Centre,  University of Sussex, Brighton, BN1 9QH, UK
}

\maketitle

\begin{abstract}
We present a general Bayesian formalism for the definition of Figures of Merit (FoMs) quantifying the scientific return of a future experiment. We introduce two new FoMs for future experiments based on their model selection capabilities, called {\em the decisiveness} of the experiment and the {\em expected strength of evidence}. We illustrate these by considering dark energy probes, and compare the relative merits of stage II, III and IV dark energy probes. We find that probes based on supernovae and on weak lensing perform rather better on model selection tasks than is indicated by their Fisher matrix FoM as defined by the Dark Energy Task Force. We argue that our ability to optimize future experiments for dark energy model selection goals is limited by our current uncertainty over the models and their parameters, which is ignored in the usual Fisher matrix forecasts. Our approach gives a more realistic assessment of the capabilities of future probes and can be applied in a variety of situations.
\end{abstract}



\section{Introduction}

As cosmology becomes increasingly dominated by results emerging from
large-scale observational programmes, it is imperative to be able to
justify that resources are being deployed as effectively as possible.
In recent years it has become standard to quantify the expected
outcome of cosmological surveys to enable comparison, a procedure
exemplified by the Figure of Merit (FoM) introduced by~\cite{Huterer_Turner} and later used by the dark energy
task force (DETF) for dark energy surveys~\citep{Albrecht:2006um,Albrecht:2009ct}. Still in its infancy,
however, is the topic of survey design, where an experiment is
optimized, within design or cost constraints, to generate the best
scientific outcome~\citep{Bassett05,BPN,Parkinson:2007cv,Parkinson:2009zi}.

Both in quantifying and in optimizing survey capability, it is important to
identify the scientific questions one hopes to answer. The DETF FoM
measures the expected parameter constraints on a two-parameter dark
energy model, using a Fisher matrix approach; this is an example of a
parameter estimation FoM, in which the correct cosmological model is
assumed to be known and the task is to estimate its parameter
values (see also e.g.~\cite{Mortonson:2010px}). However, many of the most pressing questions in cosmology
concern not parameters but models, i.e.\ the identification of the
correct set of parameters to describe our Universe. Examples are
whether cosmic acceleration is due to a cosmological constant,
quintessence, or modified gravity, and whether or not the Universe has zero
spatial curvature. These are model
selection questions, hence forecasts of the capabilities of future
probes should be assessed by their power to answer such questions, rather
than the more limited question of the error they will be able
to achieve assuming a given model is true (i.e., the usual Fisher
Matrix forecast).
Alternative FoMs, which quantify the ability of
experiments to answer model selection problems, have been previously
discussed by \cite{Mukherjee:2005tr}, \cite{Trotta:2007hy}, and \cite{BMIC_RTetal}.\footnote{For an alternative, essentially frequentist, perspective on this issue, see \cite{AmaraKitching}.}

In this paper we present a comprehensive formalism for the construction of
survey FoMs, incorporating both model and parameter uncertainty in
light of the present observational situation. In order to do so, we build on the methodology introduced in~\cite{BMIC_RTetal}. We construct two new
model selection FoMs, the {\em decisiveness} and the {\em expected
  strength of evidence}, which quantify the expected capability of an experiment
to perform model comparison tests. For illustration we focus on the
case of dark energy observations, though our formalism is broadly
applicable.

\section{Bayesian framework for performance forecasting}

\subsection{The expected utility of an experiment}

In order to build up towards the definition of our FoMs, we need to
consider the different levels of uncertainty that are relevant when
predicting the probability of a certain model selection outcome from a
future probe. Those can be summarized as follows:
\begin{itemize}
\item {\bf Level 1:} current uncertainty about the correct model
  (e.g., is it a cosmological constant or a dark energy model?). 
\item {\bf Level 2:} present-day uncertainty in the value of the
  cosmological parameters for a given model (e.g., present error on
  the dark energy equation of state parameters assuming an evolving
  dark energy model). 
\item {\bf Level 3:} realization noise, which will be present in
  future data even when assuming a model and a fiducial choice
  for its parameters.    
\end{itemize}
The commonly-used Fisher matrix forecast (see, e.g.~\cite{Tegmark:1996bz}) ignores the uncertainty arising from Levels
1 and 2, as it assumes a fiducial model (Level 1) and fiducial
parameter values (Level 2). It averages over realization noise (Level
3) in the limit of an infinite number of realizations. Furthermore, in the Fisher matrix formalism the likelihood is approximated by construction as a Gaussian, which might be inaccurate for parameter spaces exhibiting curving degeneracies and/or multimodal distributions. Clearly, the Fisher matrix procedure provides a very limited assessment of what we can expect for the scientific return of a future probe, as it ignores the uncertainty associated with the choice of model and parameter values. 

The Bayesian framework allows improvement on the usual Fisher matrix error forecast thanks to a general 
procedure which fully accounts for all three levels of uncertainty given above. This will allow us to define a new type of FoM which represents in a more realistic way the uncertainties involved in making predictions.  

Following~\cite{Loredo:2003nm}, we think of
future data $\Df$ as {\em outcomes}, which arise as consequence of our
choice of experimental parameters $e$ ({\em actions}). For each action
and each outcome, we define a utility function $\Uf(\Df, e)$. Formally, the utility only depends on the future data realization $\Df$. However, as will become clear below, the data $\Df$ are realized from a fiducial model and model parameter values. Therefore, the utility function implicitly depends on the assumed model and parameters from which the data $\Df$ are generated. The best
action is the one that maximizes the expected utility, i.e. the
utility averaged over possible outcomes:
\begin{equation} \label{def:EU}
\EU (e) \equiv \int \dr \Df p(\Df | e, \data) \Uf(\Df, e).
\end{equation}
Here, $p(\Df | e, \data) $ is the predictive distribution
for the future data, conditional on the experimental setup ($e$) and
on current data ($\data$). For a single fixed model the
predictive distribution is given by 
\be
\begin{aligned} \label{eq:predictive_general}
p(\Df | e, \data) &  = \int \dr\params \, p(\Df, \params | e, \data) \\
 			 & =  \int \dr\params \, p(\Df | \params ,  e,
 			 \data) p(\params|e,\data) \\ 
			 & = \int \dr\params \, p(\Df | \params ,  e)
 			 p(\params|\data), 
\end{aligned}
\ee
where the last line follows because $p(\Df | \params ,  e,  \data) =
p(\Df | \params ,  e)$ (conditioning on current data is irrelevant
once the parameters are given) and  $p(\params|e,\data)  = p(\params|
\data)$ (conditioning on future experimental parameters is irrelevant
for the present-day posterior). So we 
can predict the probability distribution for future data $\Df$ 
by averaging the likelihood function for the future measurement
(Level 3 uncertainty) over the current posterior on the parameters
(Level 2 uncertainty). The expected utility then becomes
\begin{equation} \label{eq:exp_utility_1}
\EU (e) = \int \dr \params p(\params| \data) \int \dr \Df p(\Df |
\params, e) \Uf(\Df, e).
\end{equation}

So far, we have tacitly assumed that only one model was being considered for the data. In practice, there will be several models that one is interested in testing (Level 1 uncertainty), and typically there is uncertainty over which one is best. This is in fact one of the main motivations for designing a new dark energy probe. If $N$ models $\{ \mdl_1, \dots, \mdl_N \}$ are being considered, each one with parameter vector $\params_i$ ($i=1,\dots, N$), the current posterior can be further extended in terms of model
averaging (Level 1), weighting each model by its current model
posterior probability, $p(\mdl_{i} | \data)$, given by
\begin{equation}
p(\mdl_{i} | \data) = \frac{p(\data| \mdl_i)p(\mdl_i)}{p(\data)},
\end{equation}
where $p(\data| \mdl_i)$ is the Bayesian evidence for model $\mdl_i$, $p(\mdl_i)$ is the model's prior and $p(\data)$ a normalizing constant. 
Using Eq.~\eqref{eq:exp_utility_1}, this gives the model-averaged expected utility
\begin{equation} \label{eq:exp_utility_2}
\begin{aligned}
\EU (e) & = \sum_{i=1}^N p(\mdl_{i}| \data) \int \dr \params_i p(\params_{i}|
\data ,\mdl_{i}) \\ & \times \int \dr \Df p(\Df | \params_{i},
e,\mdl_{i}) \Uf(\Df, e,\mdl_{i}).
\end{aligned}
\end{equation}
This expected utility is the most general definition of a FoM for a future experiment characterized by experimental parameters $e$. As we show below, the usual Fisher matrix forecast is recovered as a special case of Eq.~\eqref{eq:exp_utility_2}, as are other FoMs that have been defined in the literature, e.g.~\cite{Bassett05,Wang:2008zh,AmaraKitching}. Therefore Eq.~\eqref{eq:exp_utility_2} gives us a formalism to define in all generality the scientific return of a future experiment. This result clearly accounts for all three levels of uncertainty in making our predictions: the utility function $\Uf(\Df, e,\mdl_{i})$ (to be specified below) depends on the future data realization, $\Df$, (Level 3), which in turn is a function of the fiducial parameters value, $\params_i$, (Level 2), and is averaged over present-day model probabilities (Level 1).

\subsection{Figures of Merit from expected utility}

The expected utility of Eq.~\eqref{eq:exp_utility_2} provides the most general formalism for the evaluation of the scientific return of an experiment. It reduces to previously used FoMs for specific choices of priors and utility functions. For example, the DETF advocated using the inverse of the area of the future probe covariance matrix on the dark energy parameters as a FoM quantifying the strength of the statistical constraints from the experiment. This FoM can be recovered by setting $N=1$ in Eq.~\eqref{eq:exp_utility_2} (only one fiducial model is considered), taking a Dirac delta function for the current posterior, $p(\params|
\data ,\mdl) = \delta(\theta - \fid)$ (only the fiducial parameter vector $\fid$ is considered), assuming no realization noise (or equivalently, averaging over many future data realizations, so that $p(\Df | \params, e,\mdl) = \delta(\Df - D(\fid))$, where $D(\fid)$ describes a no-noise data realization around the fiducial parameter values, and defining the utility function as the determinant of the future Fisher matrix, evaluated at the fiducial parameter values, $\fid$. 

Another example is the Gaussian linear model considered by \cite{BMIC_RTetal}, where the utility function was chosen to be the inverse of the marginal error on the parameters of interest. It is a property of the Gaussian linear model that the error ellipse does not depend on the fiducial model nor data realization, but only on the design matrix \citep{Kunz:2006mc}. Therefore, in this case the  integration over future data $\Df$ gives unity in Eq.~\eqref{eq:exp_utility_2}, and the same expression is recovered as in \cite{BMIC_RTetal}.

\cite{Mukherjee:2005tr} defined two model selection FoMs, each of which considers two models, a cosmological constant model and a two-parameter dark energy model. One FoM asks for the strength with which the dark energy model will be excluded {\it if} the cosmological constant is correct; the current posterior is therefore taken to be that model and the FoM is the Bayes factor (defined below) in favour of the cosmological constant. The other FoM is the opposite, quantifying  whether the cosmological constant can be ruled out if the dark energy model is correct. The current posterior is now the dark energy model space, and the FoM measures in how much of that space the cosmological constant model could be excluded (for example, the inverse parameter area above a certain Bayes factor threshold, by analogy to the DETF FoM above).

\cite{Trotta:2007hy} introduced a methodology to compute the predicted posterior odds distribution (PPOD) for a model comparison from a future experiment. A PPOD-based Figure of Merit is another special case of our general formalism: it is obtained from Eq.~\eqref{eq:exp_utility_2} by assuming no realization noise, $p(\Df | \params, e,\mdl) = \delta(\Df - D(\fid))$, and adopting as utility function the tail probability of the Bayes factor obtainable by a future probe. 

For a given experimental configuration $e$, the expected utility can be evaluated as follows:
\begin{enumerate}
\item Draw a uniformly-weighted sample for the fiducial value for the parameters, $\fid$, from a Monte Carlo Markov chain distributed according to
the present, model-averaged, posterior $p(\params | \data) = \sum_{i} p(\mdl_i | \data) p(\params_i | \mdl_i, \data)$ (Levels 1 and 2). 
\item Generate pseudo-data $\Df$ for the future probe, assuming $\fid$ as fiducial parameter values. 
\item Evaluate the utility function from the future data (to be defined below).
\item Loop back to (i) and average the utility function over the so-obtained samples. 
\end{enumerate}

In general, the above procedure is computationally very expensive, as it involves two nested averages, one over the fiducial parameters (step (i)) and one over future pseudo-data realizations (step (ii)). Furthermore, in the context of model selection oriented FoMs to be introduced below, the evaluation of the utility (step (iii)) requires the computation of Bayes factors from the pseudo-data, which again is costly. If one wanted to use Markov Chain Monte Carlo (MCMC) techniques, one would typically need $\sim 10^4$ samples in step (i), and another $\sim 10^{5}$ samples to obtain a reliable estimate of the utility function in steps (ii) and (iii). Therefore, the typical number of likelihood evaluations required would be of order $\sim 10^{9}$, which is at the limit of what can be achieved today unless one adopts highly accelerated inference methods~\citep{Fendt:2006uh,Auld:2007qz,Frommert:2010zz,Bridges:2010de}. Therefore, we shall make some simplifying assumptions that reduce this computational burden very considerably. 

Firstly, we will consider only $N=2$ competing models. Secondly, we will work in the Gaussian likelihood approximation, i.e., we will assume that both the present-day and the future likelihood are well approximated by Gaussian distributions. This is the same kind of approximation involved in the usual Fisher matrix forecast. The assumption of Gaussianity further allows us to side-step the pseudo-data generation step: for a given value of the fiducial parameters, $\fid$, the maximum likelihood estimate $\hat{\params}_f$ from future data $\Df$ generated from $\fid$ is distributed as a Gaussian with mean $\fid$ and covariance matrix given by the inverse of the likelihood Fisher matrix for the future probe. As a consequence, we do not need to generate pseudo-data at all in step (ii), and we can instead work directly in parameter space, by drawing $\hat{\params}_f$ directly from a Gaussian distribution centered on $\fid$. 

Having made the above simplifications, we now turn to using the expected utility to define two new FoMs based on model selection.

\section{Figures of Merit for Model Selection}

To assess the science return of proposed missions in terms of their model selection capabilities, we propose to adopt the expected utility~of Eq.~\eqref{eq:exp_utility_2} as a
FoM for experiment $e$, after defining an appropriate utility function
$\Uf(\Df, e,\mdl_{i})$.  There are many ways to do this, and we introduce here two proposals. The first one is named {\em decisiveness}, and
it gives the probability that the proposed experiment will achieve a
decisive outcome for model selection. A good experiment should be as
decisive as possible. A complementary approach, named {\em expected
strength of evidence}, is to compute by how much the experiment is expected to
prefer one or other model on average.  Again, a good
experiment will be able to prefer one of the models strongly. 

In a two-way Bayesian model comparison, the key Bayesian statistic is the Bayes factor $B_{01}$, which
is formed from the ratio of the Bayesian evidences of the two models
being considered:
\be
B_{01} = \frac{p(\data | \mdl_0)}{p(\data | \mdl_1)},
\ee
where the Bayesian evidence is the average of the likelihood under the prior in each model:
\be
p(\data | \mdl_i) = \int \dr\params_i p(\data | \params_i, \mdl_i)p(\params_i | \mdl_i).
\ee
 The Bayes factor updates the prior probability ratio
of the models to the posterior one, indicating the extent to which the
data have modified one's original view on the relative probabilities
of the two models. The Bayes factor can be evaluated by a general numerical method such as nested
sampling \citep{Skilling,BCK,PML,MultiNest}, or, if one model is nested within the other, by the
Savage--Dickey density ratio (SDDR) \citep{Trotta:2005ar,Trotta:2008qt}. The Bayes factor is
usually interpreted on the Jeffreys' scale shown in
Table~\ref{tab:jeff} \citep{Jeff,Gordon:2007xm}.

\begin{table}
\caption{Empirical scale for evaluating the strength of evidence when
comparing two models, $\mdl_0$ versus $\mdl_1$ (Jeffreys' scale). The
rightmost column gives our convention for denoting the different
levels of evidence above these thresholds. \label{tab:jeff} }
\centering
 \begin{tabular}{l l  l} 
  $|\ln \Bf|$ & Odds & Strength of evidence \\\hline
 $<1.0$ & $\lsim 3:1$  & Inconclusive \\
 $1.0$ & $\sim 3:1$ & Weak evidence \\
 $2.5$ & $\sim 12:1$  & Moderate evidence \\
 $5.0$ & $\sim 150:1$ & Strong evidence \\
 \hline
\end{tabular}
\end{table}

\subsection{The `decisiveness' Figure of Merit}

 A `decisive' experiment is one that is able to gather strong evidence
in favour of one of the competing models. Therefore, its utility
function is 0 (1) if the Bayes factor it will obtain is below (above)
the `strong' threshold for the evidence, $\ln B = 5$, see
Table~\ref{tab:jeff} (this level of evidence is sometimes called
`decisive', hence the name of the FoM). Therefore, we are led to
the following utility function 
\be \label{eq:UF_decisive} \Uf(\Df, e,
\mdl_{i}) = \left\{
\begin{array}{rl}
1 & \textrm{~if~} |\ln\Bf| > 5 \\
0 & \textrm{~otherwise,} 
\end{array} \right. 
\ee
where  $\Bf$ is the Bayes factor between the two models, obtained by the future experiment $e$. The best experiment is the one that maximizes
this quantity, i.e.\ the one whose probability of obtaining a strong
model selection outcome for either of the models is maximized. We thus define the {\em
decisiveness} $\Dec$ of an experiment $e$ as its expected utility, Eq.~\eqref{eq:exp_utility_2}, with the utility function \eqref{eq:UF_decisive}.
We note that $\Dec$ as a Figure of Merit is especially resilient to the 
scatter in the Bayes factor coming from averaging over data realizations and the unknown fiducial parameter values
\citep{Jenkins:2011va}. In fact, our formalism takes this scatter
into full account, and if too many realizations are scattered out of the 
`decisive' region (e.g. due to large noise on the measurements from the future probe) then this will lead to a lower Figure of Merit. Therefore, using $\Dec$ to optimize the design of an experiment is particularly useful to guard against this effect.

\subsection{The `expected strength of evidence' Figure of Merit} 

Instead of the discrete utility function above, we can adopt
one that is more gradual in assessing the merit of the future
probe. Such a utility function is  
\be
 \label{eq:UF_expected_strength_of_evidence} \Uf(\Df, e, \mdl_{i}) = (-1)^i \ln\Bf,
 \ee 
 which describes the strength of the model selection result from the future probe. By plugging this utility function into 
Eq.~\eqref{eq:exp_utility_2}, we obtain a FoM that we call the
`expected strength of evidence' and denote by $\Expu$. The rationale is
that for every given fiducial value of the parameters and for every
data realization, the best experiment is the one that maximizes the
support to the true model (i.e., the model out of which the data
actually come from), even though it might be that the experiment in
question is not strong enough to achieve decisiveness. 
 
The factor $(-1)^i$ in Eq.~(\ref{eq:UF_expected_strength_of_evidence}) is to ensure that the utility only rewards support for the {\em correct} model; e.g.\ under the more
complex model ($\mdl_{1}$), we want to maximize $-\ln\Bf$, the
odds in favour of $\mdl_{1}$. Bayes factors can occasionally favour the wrong model, e.g.\ if the true model were a dark energy model with $w=-0.999$, anything other than an extraordinarily precise experiment is likely to favour the more predictive cosmological constant model. Nevertheless, support for the wrong model will happen only in a small parameter space region and will be overwhelmed when the average over the current posterior is carried out, making the above nearly equivalent to the simpler choice $\Uf(\Df, e, \mdl_{i}) = |\ln\Bf|$. We have found in the dark energy application presented below that for all future dark energy probes the difference in the FoM between these two choices is less than about 5\%, so in practice almost negligible. 

It might seem at first glance that an experiment that maximizes the expected strength of evidence is also one that minimizes the error ellipse in the parameter space of interest. If this was true, than the ranking of probes obtained with the expected strength of evidence would be the same as the one from the DETF FoM. However, consider the
SDDR expression for nested models~\citep{Trotta:2005ar}: 
\be \label{eq:SDDR}
\Bf = \frac{p(\phi | \Df, e, \data, \mdl_{1})}{p(\phi | \mdl_{1})}
    {\Large\vert}_{\phi=\phi_{0}} ,
\ee
where $\phi$ are the extra parameters of interest for the more
complicated model, which reduces to the simpler model for
$\phi=\phi_{0}$. The odds against $\mdl_{0}$ are maximized when the
marginal posterior on the extra parameters is as small as possible at
the location in parameter space predicted by the simpler model. This
means that maximizing $-\ln\Bf$ requires minimizing the posterior
error {\em along the direction connecting the fiducial value of
  $(w_{0}, w_{a})$ to $(-1,0)$} (if we restrict our consideration to
the dark energy example, where $\phi=(w_{0},w_{a})$). In other words,
the expected strength of evidence FoM favours experiments that deliver error
ellipses whose most tightly constrained principal direction points
towards the location of the simpler model in parameter space, hence
minimizing model confusion. If instead the data come from $\mdl_{0}$, then
the utility function requires that the height of the posterior at the
location of the true model be as large as possible. Since the
posterior is normalized, this requires the posterior to be as tightly
constrained around the true value as possible, which is obviously
desirable.

To summarize, the decisiveness FoM $0 \leq \Dec \leq 1$ can be understood as an absolute scale
measuring the model selection capabilities of an experiment, with
$\Dec = 1$
denoting the maximum possible performance in terms of model comparison
utility (i.e., an experiment that is guaranteed to achieve a decisive
model selection result). On the other hand, many probes might still be
interesting to build but may fall short of the achieving strong
evidence anywhere in parameter space, hence such experiments would all
have $\Dec = 0$. Yet it is still a relevant question to try and rank
them according to their merits. This can be done by looking at the
expected strength of evidence, which always returns a non-zero
value. Therefore, the expected strength of evidence $\Expu$ can be regarded as
a relative scale of the capabilities of the probes.


\section{Application to future dark energy probes}

We now apply our newly defined model selection FoMs to a set of representative proposals for future dark energy probes. We consider a $\Lambda$CDM model with dark energy in the form of a cosmological constant versus an evolving dark energy model where the equation of state is $w(z) = w_0 + w_a z/(1+z)$, described by the two parameters $(w_0, w_a$). This is a case of nested models, i.e., where the simpler model (the cosmological constant) is obtained as a special case of the evolving dark energy model by setting $w_0 = -1, w_a = 0$. The other cosmological parameters (common to both models) are the baryonic density, the dark matter density, the spatial curvature, the amplitude of scalar adiabatic fluctuations and the spectral index of perturbations. We include curvature in our analysis as this impacts strongly on the constraints on evolving dark energy models~\citep{Wang:2007mza,Clarkson:2007bc}.

The current posterior is obtained using the following data sets: WMAP5 \citep{Dunkley:2008ie}, Acbar07 \citep{Kuo:2006ya}, CBI \citep{Sievers:2005gj}, BOOMERANG03 \citep{Jones:2005yb} for the CMB, SDSS LRG DR4 \citep{Tegmark:2006az} for $P(k)$, the Hubble Key Project determination of $H_0$ \citep{Freedman:2000cf}, big bang nucleosynthesis limits on $\Omega_b h^2$ \citep{Kirkman:2003uv}, and the Union supernova-Ia compilation \citep{Kowalski:2008ez}. The priors on the common parameters are irrelevant as they cancel from the Bayes factor between the two models (as long as those priors are sufficiently wide to include the maximum likelihood and uncorrelated with the dark energy priors, see~\cite{Trotta:2005ar}), so the only important prior is the one on $(w_0, w_a)$. We choose a Gaussian prior centered on $w_0 = -1, w_a = 0$ with Fisher matrix $\Pi = {\rm diag}(1, 1/2)$. With this prior and the above data sets, we obtain a Bayes factor $B_{01} = 13.7$ in favour of the $\Lambda$CDM model (representing moderate evidence against an evolving dark energy). This means that 93\% of samples from the current posterior will be drawn from a $\Lambda$CDM model, and 7\% from a model with evolving dark energy.

\subsection{Future dark energy probes}

We use a selection of future missions based on the DETF classification \citep{Albrecht:2006um}, using Fisher matrices provided by the DETFast package \citep{Dick:2006zzz}. This package provides only the Fisher matrices evaluated at a fixed fiducial $\Lambda$CDM cosmology, so we have to assume that the Fisher matrices do not vary significantly for different fiducial parameters drawn from the current posterior. In other words, we take the Fisher matrix for the future experiment at a fiducial $\Lambda$CDM point and translate it in parameter space, without recomputing it for each new sample of $\fid$. This is clearly an oversimplification, but since the dark energy parameters are the most important ones for this application, and since 93\% of points drawn from the current posterior belong to the $\Lambda$CDM case, we expect that the results are not too strongly biased. We intend to study the impact of this assumption and to provide a more comprehensive study of the power of future dark energy probes in future work, while using the simplified approach as an illustration of our new FoMs here.

The Dark Energy Task Force has classified the dark energy probes in stages, with stage II being those that are currently ongoing or completed, stage III being medium-term projects and stage IV future large projects (optical large survey telescopes, `LST', space-based missions, `S', and the square kilometer array, `SKA'). The probes that we consider here include weak lensing (WL), type-Ia supernovae (SN), Baryon Acoustic Oscillations (BAO), cluster counts (CL) and combinations of several probes (ALL). A suffix `-o' and `-p' denotes optimistic and pessimistic assumptions about systematic errors. The `p' in the names of the stage III experiments signals the use of photometric redshifts while an `s' is used for spectroscopic surveys (that tend to cover a much smaller area). For further, detailed information please consult the DETF report.

The utility function computation proceeds as follows. In order to evaluate the decisiveness, Eq.~\eqref{eq:UF_decisive}, and expected strength of evidence, Eq.~\eqref{eq:UF_expected_strength_of_evidence}, we need the Bayes factor $\ln\Bf$ for the future experiment. This is obtained analytically via the SDDR formula, Eq.~\eqref{eq:SDDR}:
\be \label{eq:B01future}
 \ln \Bf = \frac{1}{2}\ln \frac{|\Pi|}{|F_\phi|} - \frac{1}{2}(\bar{\phi} - \phi_0)^t
 F_\phi (\bar{\phi} - \phi_0) 
 \ee
 where $\phi = (w_0, w_a)$ are the dark energy parameters of interest, $\Pi$ is their prior Fisher matrix and $F_\phi$ is the marginal posterior Fisher matrix for $\phi$. We have defined $\phi_0 = (-1, 0)$ and $\bar{\phi}$ is the posterior mean from both current and future data. This can be obtained as the $\phi$-components of the posterior mean vector in the full parameter space,  
 \be
 \bar{\params} = F^{-1}(L^f \hat{\params}_f + L \hat{\params} + \Pi\theta_0).
 \ee
 In the above, $L^f$ is the future probe likelihood Fisher matrix, $L$ is the current constraints Fisher matrix, $\theta_0$ is the prior mean, $\hat{\params}_f$ is the future maximum likelihood location while $\hat{\params}$ is the present constraints' maximum likelihood point. The Fisher matrix from the future and present data, $F$, is given by
 \be
 F = L^f + L + \Pi.
 \ee
 
The prior used in Eq.~\eqref{eq:B01future} is the same as the one adopted for the analysis of the present-day data. This is because the prior in the context of Bayesian model selection should be understood as representing the {\em a priori} plausible parameter values under the model. Therefore, we do not update the prior to the posterior from the present-day inference step when evaluating the future Bayes factor. The likelihood is obtained from the Fisher matrix formalism, with the above-mentioned additional assumption that the future likelihood Fisher matrix is independent of the fiducial parameter value adopted.

\begin{table} 
\caption{Results for FoMs of various dark energy probes. $\Dec$ is the decisiveness given in Eq.~(\ref{eq:UF_decisive}) and $\Expu$ is the expected strength of evidence, Eq.~(\ref{eq:UF_expected_strength_of_evidence}).
\label{table:results}}
\centering
 \begin{tabular}{l | l l l } 
 Experiment &        DETF    FoM&    $\Dec $&   $\Expu$  \\ \hline
CL-II	 &	0.13 &		0	&	2.3 \\
SN-II	 &	$1.4\times10^{-2} $	& $2.0\times10^{-3} $ & 2.7 \\
WL-II &                   0.7    &     $4.3\times10^{-3} $         &         2.8 \\
\hline
BAO-IIIp-p	 & $7.1\times10^{-5} $ & $2.0\times10^{-4} $ & 2.5 \\
BAO-IIIs-p	 & 0.87		& $1.2\times10^{-3} $ & 2.7 \\
BAO-IIIs-o	 & 1.0		& $1.5\times10^{-3} $ & 2.7 \\
CL-IIIp-p &	0.56	& $2.9\times10^{-3} $ &	2.7 \\
CL-IIIp-o &	8.5	& $1.5\times10^{-2} $ &	3.4 \\
SN-IIIp-p	& $4.9\times10^{-3} $ & $1.5\times10^{-3} $ &	2.7 \\
SN-IIIp-o	& $2.0\times10^{-3} $ & $8.5\times10^{-3} $ &	3.1 \\
SN-IIIs	& $4.2\times10^{-2} $ & $5.9\times10^{-3} $ &	2.9 \\
WL-IIIp-p    &           6.4      &   $9.9\times10^{-3} $          &         3.2 \\
WL-IIIp-o    &          17      &  $1.6\times10^{-2} $          &         3.5 \\
ALL-IIIp-p	& 59	&	$2.7\times10^{-2} $ 	&	4.1 \\
ALL-IIIp-o	& 150	&	0.53	&	5.1 \\
ALL-IIIs-p	& 130	&	0.38	&	4.8 \\
ALL-IIIs-o	& 200	&	0.58	&	5.3 \\
\hline
BAO-IVLST-p	& $1.8\times10^{-2} $ 	&$1.2\times10^{-3} $ 	&	2.6 \\
BAO-IVLST-o	& $4.0\times10^{-2} $ 	&$1.6\times10^{-3} $ 	&	2.7 \\
BAO-IVSKA-p	& 1.3		& $4.3\times10^{-3} $ 	&	3.0 \\
BAO-IVSKA-o	& 3.4		& $9.0\times10^{-3} $ 	&	3.3 \\
BAO-IVS-p	& 1.4		& $3.7\times10^{-3} $ 	&	3.0 \\
BAO-IVS-o	& 3.4		& $7.0\times10^{-3} $ 	&	3.2 \\
CL-IVS-p	& 0.50	& $3.1\times10^{-3} $ 	&	2.8 \\
CL-IVS-o	& 9.5		& $1.6\times10^{-2} $ 	&	3.5 \\
SN-IVLST-o	& 0.32	& $1.4\times10^{-2} $ 	&	3.4 \\
SN-IVS-p	  & 0.65		& $9.9\times10^{-3} $ 	&	3.2 \\
SN-IVS-o	  & 0.76		& $1.6\times10^{-2} $ 	&	3.5 \\
WL-IVLST-p   &          15    &   $1.3\times10^{-2} $          &         3.5 \\
WL-IVLST-o   &         170  &       0.77            &         6.0 \\
WL-IVSKA-p  &            4.6    &     $1.9\times10^{-2} $        &           3.6 \\
WL-IVSKA-o  &          280      &   0.81        &             6.3 \\
WL-IVS-p  &             83      &    0.37           &          4.7 \\
WL-IVS-o  &            140      &   0.67           &          5.5  \\
ALL-LST-p	&   180		&   0.54		&   5.1 \\
ALL-LST-o	&   900		&   0.89		&   6.9 \\
ALL-SKA-p	&   160		&   0.49		&   5.0 \\
ALL-SKA-o	&   950		&   0.90		&   7.0 \\
ALL-IVS-p		&   480		&   0.81		&   6.2 \\
ALL-IVS-o		&   900		&   0.90		&   6.9 \\
\hline
\end{tabular}
\end{table}

\begin{figure*}
\includegraphics[width=0.95\linewidth]{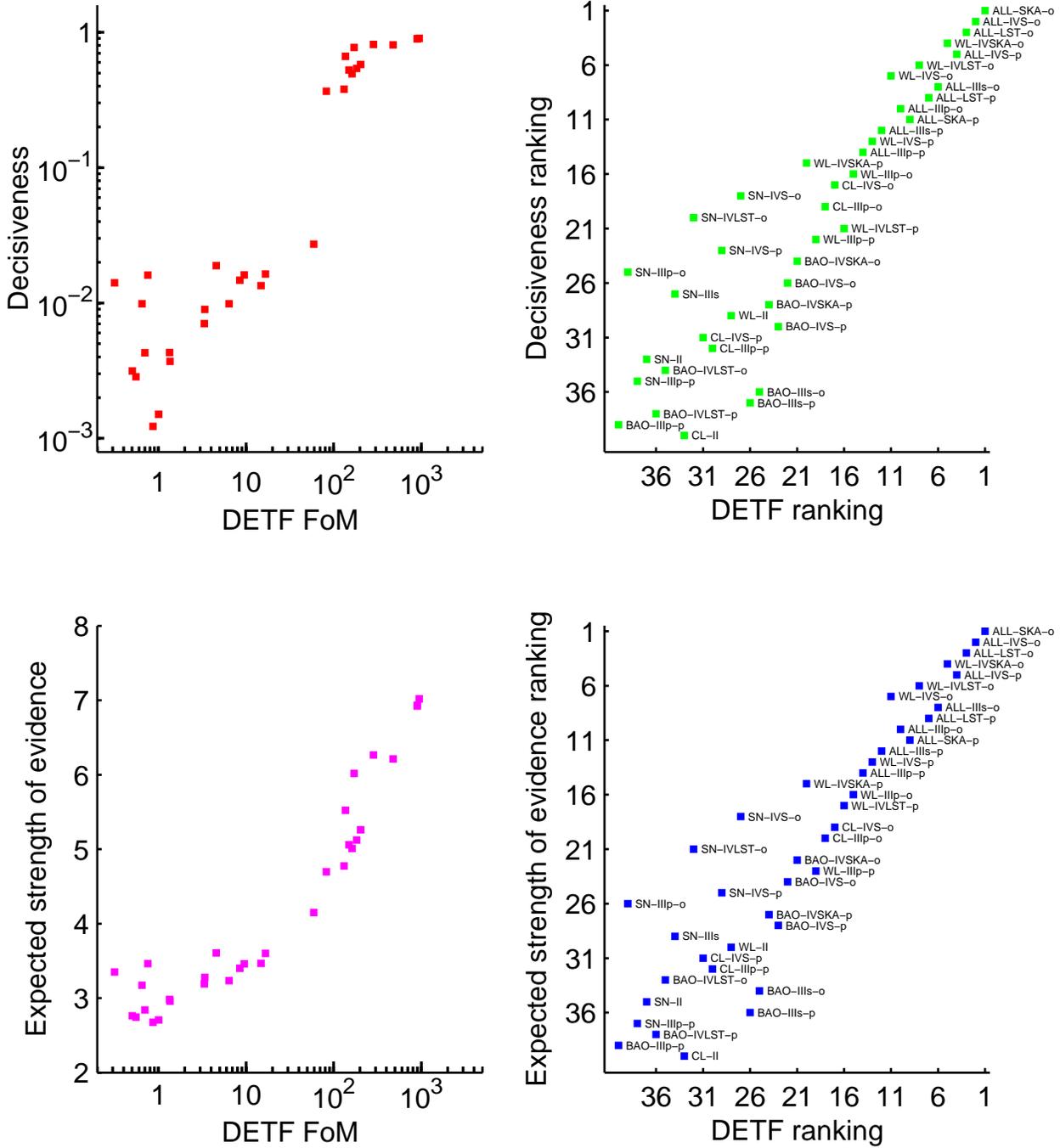}
\caption{Comparison of our model selection FoMs to the DETF FoM (left panels) and the ranking of dark energy probes derived from them (right panels). \label{fig:FoM}}
\end{figure*}

Some of the dark energy probes can achieve a very strong model selection in favour of an evolving dark energy model in parts of the parameter space, often obtaining $\ln\Bf \ll -100$. This would correspond to a detection of a non-constant equation of state at many sigma confidence level. However, we do not expect our Gaussian approximation to the likelihood to hold true so far into the tails of the distribution. Therefore, in order to be conservative, we impose a floor at $\ln\Bf = -20$ when computing the expected strength of evidence from Eq.~\eqref{eq:UF_expected_strength_of_evidence}: any value of $\ln\Bf$ below this floor is remapped to the floor value.

\subsection{Results}

The results for the future probes are presented in Table~\ref{table:results} and plotted in Fig.~\ref{fig:FoM}, where we compare the DETF FoM with our new model selection FoMs. We notice that the decisiveness FoM separates the sample into two distinct groups, those with $\Dec \lesssim 0.1$ (single probes up to level III and several pessimistic single probes at level IV, together with BAO-IVS-o) that are unlikely to provide a decisive answer to the question whether dark energy is dynamical or not, and the rest with $\Dec \gsim 0.1$. This division is not apparent in $\Expu$ and the DETF FoM, and it leads to critical values of $\Expu \approx 4$ and DETF FoM $\approx 70$ below which an experiment is unlikely to obtain a strong model selection result.

The ranking of the experiments between $\Dec$ and $\Expu$ is almost the same, while the DETF FoM gives a similar but not always identical ranking. Looking at the right-hand panels in Fig.~\ref{fig:FoM}, we notice that the WL and SN probes tend to lie above the trend line (are more likely to provide a decisive model selection result than would be expected from the DETF FoM) while spectroscopic BAO probes lie below.
 
\begin{figure}
\includegraphics[width=90mm]{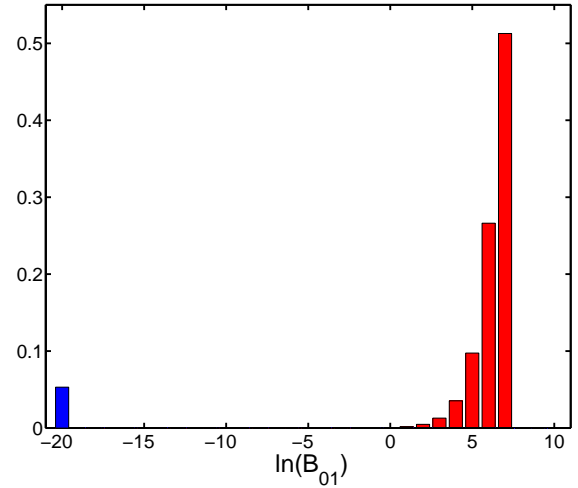}
\caption{Histogram of $\ln(\Bf)$ values for the ALL-SKA-o DETF case. Red bars (those on the right) show cases drawn from a $\Lambda$CDM model (93\% of cases according to current posterior) and the blue bar those with an evolving dark energy fiducial model (7\% of cases, capped at $\ln(\Bf)=-20$ as described in the text). \label{fig:bdist}}
\end{figure}

In Figure \ref{fig:bdist} we show the distribution of $\ln(B_{01})$ for $10^6$ outcomes for the ALL-SKA-o probe (the most powerful probe considered here). The red bars on the right hand side are for data drawn from a $\Lambda$CDM model, for which this probe often but not always achieves a decisive outcome. The blue bar on the left shows that the probe will deliver very powerful results if the dark energy is actually evolving (given the priors adopted, and current knowledge on dark energy parameters). It is not surprising that the model selection outcomes against $\Lambda$CDM tend be stronger than those that support it: it is always more difficult to strongly support a nested model, as the simpler model only ``profits'' from its predictiveness (thanks to the Occam's razor effect), but can never provide a better fit.
 

\section{Conclusions}

We have presented a general Bayesian formalism for the definition of FoMs encapsulating the expected scientific return of a future experiments. Our method fully accounts for all source of uncertainties involved in the prediction, including present-day model and parameter uncertainties, and realization noise. It thus improves on the usual Fisher matrix methods by producing more realistic forecasts for the possible distribution of future experimental outcomes. 

We used this framework to define two Figures of Merit for probes that measure the dark energy equation of state in order to test the $\Lambda$CDM paradigm: the {\em decisiveness} $\Dec$ which quantifies the probability that a probe will deliver a decisive result in favour or against the cosmological constant, and the {\em expected strength of evidence} $\Expu$ that returns a measure of the expected power of a probe for model selection. We compared these quantities to the widely-used DETF FoM for a range of probes, and found that the rankings agree reasonably well, but that weak lensing and supernova probes have a higher than expected model selection power relative to their DETF FoM ranking. We also found, for our choice of prior, that there is a critical DETF FoM of around 70 below which probes are very unlikely to obtain a strong model selection result.

An additional advantage of the formalism presented in this paper, and of any Figures of Merit that use it, is the possibility to include further observations, for example those that constrain the growth history or the presence of effective anisotropic stresses. One just extends the likelihood based on the predictions of the underlying models, but the procedure is unchanged, and the interpretation of the results is unchanged as well. There is therefore no need to define new FoM's as data analysis goals for future probes evolve.

The methodology presented here is widely applicable to a variety of forecasting and optimization problems. Our application to the model selection capabilities of future dark energy missions is but a first step towards a fully Bayesian approach to performance forecast.


\section*{Acknowledgments} 
R.T. was partially supported by the
Royal Astronomical Society through the Sir Norman Lockyer Fellowship
and by St Anne's College, Oxford. M.K.\ and A.R.L.\ acknowledge support from the Science and Technology Facilities Council [grant number ST/F002858/1], and M.K.\ from the Swiss NSF.

\end{document}